\documentclass{article}
\usepackage{bezier,emlines2,gdgspace}
\advance\textheight by -\headsep\advance\textheight by -\headheight
\begin{document}
\large

\centerline {\Large Weak Values and Consistent Histories in Quantum Theory}
\centerline{December 5, 2002}
\centerline{R. E. Kastner}
\centerline{Department of Philosophy}
\centerline{University of Maryland}
\centerline{College Park, MD 20742}
\vskip 1cm
ABSTRACT: A relation is obtained between weak values of quantum observables
and the consistency criterion for histories of quantum events. It is shown
that ``strange'' weak values for projection operators (such as values less than zero) 
always correspond to inconsistent families of histories. It is argued
that using the ABL rule to obtain probabilities for counterfactual
measurements corresponding to
those strange weak values gives inconsistent results. This problem is shown to
be remedied by using the conditional weight, or pseudo-probability,
obtained from the multiple-time application of Luders' Rule.
It is argued that an assumption of reverse causality
(a form of time symmetry) implies that
weak values obtain, in a restricted sense, at the time of the weak measurement as well
as at the time of post-selection. Finally, it is argued that weak values
are more appropriately characterised as multiple-time amplitudes than expectation values, 
and as such can have little to say about counterfactual questions. 

\newpage
1. Introduction

The ``weak value'' of a quantum mechanical observable, a concept first introduced by
Aharonov and Vaidman (1990), is a generalized 
``expectation value''\footnote{\normalsize ``Expectation values'' is
put in scare quotes here since weak values differ in important ways
from the usual expectation value; see section 7.} for the case of pre- and post-selection. 
A weak value of the operator
$O$ with respect to states $|a\rangle$ and $|b\rangle$ is
defined as:

$${\langle O\rangle}_w  =   {\langle b|O|a\rangle 
  \over  \langle b|a\rangle}\eqno(1)$$

Weak values get their name from the
fact that they can often only be measured through a ``weakened''
 measurement procedure in which the interaction Hamiltonian 
provides
only a slight coupling between apparatus and system. This results in a significant
 imprecision
which requires the measurement of a large number of individual systems
---i.e., an
ensemble---in order to measure a single weak value. It should be noted, however,
that the term ``weak value'' is perhaps something of a misnomer, since it is simply
defined---independently of measurement---as the normalized inner product of two states,
one of which is acted upon by a Hermitian operator. So-called ``weak values''
{\it can} take on ``sharp'' values, that is, eigenvalues of the given
operator. 
The nature of the weak value depends
crucially on the entire context of operator and the two chosen states.
It is this dependence which is the primary subject of this paper.

 For clarity, let us define the following terms for the various types
of weak value:

1. ``Sharp'' weak value (SWV): a weak value that coincides with
an eigenvalue of the operator.

2. ``Unsharp'' weak value (UWV): a weak value that lies within
the range of eigenvalues of the operator but which is not
equal to any eigenvalue of the operator.

3. ``Strange'' weak value (STWV): a weak value that lies outside
the range of eigenvalues of the operator.

Weak values obey additivity, that is:

$$ {\langle A + B\rangle}_w = {\langle A \rangle}_w + {\langle B \rangle}_w, 
  \eqno(2)$$

In addition, the discussion in this paper is restricted
to the weak values of projection operators only,
since these are the ones that occur in the context of
histories of events.

2. Weak Values and the Consistency Criterion for Event Histories

As alluded to at the end of the previous section, when the operator
$O$ in (1) is a projection operator, a weak value  
effectively defines a ``history,'' i.e.,
a particular sequence of events. That is, in terms of a pre-
and post-selection experiment, the weak value ${\langle O\rangle}_w$ describes
the following sequence of events: a system is prepared in
state $|a\rangle$ at time $t_0$, possesses the property associated
with $O = |o\rangle\langle o|$ at time $t_1$,
and is post-selected in state $|b\rangle$ at $t_2$. Therefore one way to learn more about weak
values is to study them in terms of Griffiths' formulation of consistent
histories (cf. Griffiths 1996--2002). In Griffith's terms, a ``family'' of histories
is a set of different possible sequences of events. Such a family
can be consistent or inconsistent, depending on whether one can
construct a standard (``classical'') probability space accomodating all the histories.
This will be described in more detail in what follows.

Consider a family of histories comprised of two
possible sequences of events $Y$ and $Y'$, where the events occur
at times $t_0, t_1$ and $t_2$ consecutively:

$$Y = \{D \to E \to F\}  \eqno(3) $$ 
and
$$Y'= \{ D\to  E'\to   F\}\eqno(3')$$

In the above expressions, D is the projection operator corresponding to the preselection state 
$|D\rangle$, F is the
projection operator corresponding to the post-selection state $|F\rangle$ , and E and E'
are complementary, exhaustive projections  of a complete observable 
$\cal E$ possibly measured
between events D and F. In other words, $E' = (1-E)$.

If the family is consistent, that means that the probabilities of the
two histories are additive, i.e.:

\hskip 1.5cm Prob$(Y \vee Y') =$ Prob$(Y)\  + $ Prob$(Y')$. 
 \hskip 5cm (4)\vskip .2cm

(where $(Y \vee Y')$ denotes the disjunction of the two histories).

 Loosely speaking, (4) requires
that the two histories do not ``interfere'' with
each other. The two-slit experiment famously violates this condition,
if we think of Y and Y' as representing the particle leaving the source,
going through one slit or the other, respectively, and landing on the screen.
 
The consistency condition as given by Griffiths (1996) is concisely written
as:\footnote{\normalsize Using the Schrodinger 
operator representation and zero Hamiltonian between the events.}

$$Tr [F E D E' ] = 0.		\eqno(5)$$

The definition of weak value of an observable E in the context of such
a history is

$${\langle E\rangle}_w  =   {\langle F|E|D\rangle   \over  
\langle F|D\rangle}\eqno(6)$$

The consistency condition (5) can be written in terms of weak values
 (6) as follows:

$$Tr [ F E D E' ]  = Tr [ |F\rangle \langle F|E|D\rangle \langle D|E'\rangle 
 \langle E'|] $$ 

$$=  \langle F|D\rangle {\langle E\rangle}_w  Tr \{|F\rangle  \langle D|E'\rangle 
 \langle E'|\}$$ 

$$=  \langle F|D\rangle {\langle E\rangle}_w  \langle D|E'\rangle 
 \langle E'|F\rangle $$      

$$= |\langle F|D\rangle  |^2  {\langle E\rangle}_w 
 {\langle E'\rangle}^*_w\}  = 0	\eqno(5')$$

Equation (5') expresses the consistency condition in terms of the weak values for the
alternative possible events occurring between D and F.

Now, suppose we are interested in the weak value of E for a pre- and post-selection
experiment. The normal eigenvalues for projection operators are zero or one;
what does the consistency condition (6) tell us for such normal values?

Recall first that  $\langle F|D\rangle$   is always nonzero, otherwise the weak value 
is undefined.
If ${\langle E\rangle}_w = 1$, then ${\langle E'\rangle}_w =
 1 - {\langle E\rangle}_w 
= 0$, by the additivity of weak values (see equation 23); so (5') is satisfied. On the other hand, if 
${\langle E\rangle}_w = 0$, then (5') is also satisfied.

If, however, the weak value of E is neither zero nor 1, clearly
the right side of equation (5') will not vanish
and the consistency condition will not be satisfied.

There are two different ways in which (5) can fail. The first, when the right
hand side is nonzero but positive and less than 1, corresponds to the 
Unsharp situation (UWV) in which
the weak values do not coincide with allowed eigenvalues but are still
within the allowed range of eigenvalues. The second, in which the right hand
side is complex or negative, corresponds to situations with ``strange'' 
weak values (STWV)---that
is, values outside the range of eigenvalues.

\vskip .5cm

3. Example: The Three-Box Experiment

The ``strange'' weak values referred to above 
appear in the 3-Box experiment (cf. Vaidman 1996,1999; Griffiths 1996, 2002; Kastner 
1999a,b) as well as the 
Hardy (1992) example of two overlapping interferometers as analyzed by Aharonov et al
(2001) (see Figure 1). In the 3-Box experiment, which has been discussed
extensively in the literature (see references above), an apparent paradox is
obtained in which an appropriately pre- and post-selected system seems
to be in ``two boxes at once'' with certainty. The paradox uses pre-
and post-selected systems with the following pre- and post-selection
states:

\def\postone{|\phi\rangle}
\def\posttwo{|\phi \prime\rangle}
\def\postthree{|\phi \prime\prime\rangle}
\def\statea{\vert a\rangle}
\def\stateb{\vert b\rangle}
\def\statec{\vert c\rangle}
\def\stateapr{\vert a'\rangle}
\def\statebpr{\vert b'\rangle}

Particles are pre-selected in state 
$$|\psi\rangle = {1\over \sqrt 3}
 \biggl(\statea + \stateb + \statec\biggr)\eqno(7a)$$

and post-selected in the state

$$|\phi\rangle = {1\over \sqrt 3} \biggl(\statea + \stateb 
- \statec\biggr),\eqno(7b)$$

where $\statea$, $\stateb$, and $\statec$ correspond to each of the
three boxes and form an orthonormal basis for the system's Hilbert space.

Now, it turns out that when one calculates the probability of such a system
being in Box A (observable $|a\rangle\langle a|$) at a time between pre- and post-selection, one
finds that the system is in Box A with certainty; but doing the same
calculation for Box B (observable $|b\rangle\langle b|$) also yields a value of certainty. Thus
it appears that the system is somehow in ``two boxes at 
once.''\footnote{\normalsize The calculation in question is done
using the ABL rule, discussed in some detail in section 6.}

Vaidman (1996) also shows that the weak
value of the observable corresponding to whether a particle is in the third box, for the
given pre- and post-selected states, is $-1$, a ``strange'' weak value. 
This result can be obtained either through a direct calculation of
(1) for the $C$ observable or through additivity (2), using
the fact that $C = 1 - (A + B)$,
and noting that ${\langle A\rangle}_w = {\langle B\rangle}_w = 1$.

Similarly, in the Hardy example, 
the weak value
of the observable corresponding to having an electron-positron pair in the
non-overlapping arms of the interferometers for the given pre- and post-selection 
is $-1$. In what follows
we analyze each of these examples. For the 3-box experiment we consider
 an explicit post-selection
observable basis to see more clearly how the weak value manifests itself
as an apparatus state, and how a time symmetry postulate affects the interpretation
of that state. In the Hardy example, it is verified that the ``strange'' weak value
corresponds to a characteristic failure of the consistency condition
as discussed in section 2.

First, consider the three-box experiment
for the case in which the observable corresponding to opening box C
is weakly measured at time $t_1$ (with an associated weak value of $-1$).

We also define an orthonormal ``post-selection'' basis
containing the post-selected state $\postone$: let this be
the set

$$\postone = {1\over \sqrt 3} \biggl(\statea + \stateb - \statec\biggr)\eqno(8a)$$

$$\posttwo = {1\over \sqrt 2} \biggl(\statea - \stateb \biggr)\eqno(8b)$$

$$\postthree = {1\over \sqrt 6} \biggl(\statea + \stateb + 2\statec\biggr)\eqno(8c)$$

The apparatus is in the initial unsharp ready state (as projected
onto the pointer variable $x$ basis): 

$$\langle x|\chi _0 \rangle =  \bigl({1\over \pi^{1\over 4} \Delta^{1\over 2}}\bigr)\ 
exp\Bigl({(x-{x_0})^2\over 2\Delta^2}\Bigr)       \eqno(9)$$

where the uncertainty $\Delta$ in pointer position is much larger than
the difference between pointer positions corresponding to
measurement results, and $x_0$ is the ready position.

The apparatus pointer states for measurement results (in this
case for projection operators) are 

$$\langle x|\chi _1 \rangle=  \bigl({1\over \pi^{1\over 4} \Delta^{1\over 2}}\bigr)\ 
exp\Bigl({(x-1)^2\over 2\Delta^2}\Bigr) \eqno(10a) $$ 

corresponding to $1$ or ``yes,'' and

$$\langle x|\chi _2 \rangle= \bigl({1\over \pi^{1\over 4} \Delta^{1\over 2}}\bigr)\ 
exp\Bigl({x^2\over 2\Delta^2}\Bigr)  \eqno(10b)$$

corresponding to $0$ or ``no.''

 The sequence
of events is described as follows:

At $t_0$ the combined state for system + apparatus is:

$$|\Psi_0 \rangle = |\chi_0 \rangle  \otimes |\psi\rangle. \eqno(11)$$

At the intermediate time $t_1$ the interaction Hamiltonian
establishes a correlation yielding the entangled state:

$$|\Psi_1 \rangle = {1\over \sqrt 3} |\chi _1 \rangle \otimes |c\rangle + {1\over \sqrt 3} |\chi _2 \rangle \otimes 
\biggl(|a\rangle + |b\rangle\biggr) \eqno(12)$$

Rewriting this in the post-selection basis, we find:

$$|\Psi_1 \rangle = |\chi _1 \rangle \otimes \Biggl( -{1\over 3} \postone + {\sqrt 2\over 3} \postthree\Biggr)
+ |\chi _2 \rangle \otimes \Biggl( {2\over 3} \postone + {\sqrt 2\over 3}
 \postthree\Biggr)\eqno(13)$$

If we now collect terms in the system post-selection basis, an apparatus superpositon
becomes apparent:

$$|\Psi_1 \rangle = \Biggl( -{1\over 3} |\chi _1 \rangle + {2\over 3} |\chi _2 \rangle \Biggr) \otimes \postone
+ {\sqrt 2\over 3} \Biggl( |\chi _1 \rangle+|\chi _2 \rangle\Biggr)
 \otimes \postthree \eqno(14)$$

All of the above, (12) through (14), apply to the same intervening time $t_1$.
At time $t_2$, postselection of the state $\postone$ occurs and only the
first term remains; thus we obtain finally

$$|\Psi_2) = \Biggl( -{1\over 3} |\chi _1 \rangle + {2\over 3} |\chi _2 \rangle \Biggr) 
\otimes \postone \eqno(15)$$

Before going further, it should be noted that the term ``measurement'' 
as applied to the process at $t_1$ is somewhat inaccurate for the following
reason. At time $t_1$, all that has happened is that a correlation has been
established between the apparatus states and eigenstates of C; 
no single term of the resulting
superposition has yet been ``projected out'' in the sense of a
state vector collapse or von Neumann projection postulate.\footnote{\normalsize
Cf. von Neumann (1932), p. 553.}
 Therefore, to
avoid confusion, in what follows I will refer to the process at $t_1$ as a ``partial measurement.''
However, it should be noted that one {\it can} complete the measurement on the
apparatus only (i.e., record a definite pointer result) without
disturbing the system, since the collapse will only occur in the
apparatus Hilbert space---i.e., the collapse will be with respect to the sharp
pointer operator $X$
on the apparatus Hilbert space only. Discussions of weak values usually assume
only a partial measurement at $t_1$ because it is simpler to analyze.

The above steps can be considered as equally applying to the case
of a sharp partial measurement of C at time $t_1$ (in which case
the apparatus states, corresponding to the ``which box'' operator
(in this case $C$), are orthogonal, or have zero spread).
We see that partially measuring C, either sharply or weakly,
 creates a superposition of apparatus states
corresponding to the ``normal'' values of ``yes'' or ``no'' 
for the presence of the particle in whichever box is being opened.
This indeterminacy of the
apparatus state with respect to the post-selection state is characteristic
of an inconsistent family of histories (in this case, the two
possible outcomes of ``in box C'' or ``not in box C'', given the
pre- and post-selection states). 

For the unsharp observable states $|\chi _i \rangle$, the mean value of the pointer location
$x$ turns out to be $-1$, which can be measured through a statistical analysis
of a large number of identically pre- and post-selected systems.

4.  The Hardy Experiment

As observed by Aharonov et al (2001), the Hardy experiment 
constitutes another example of a ``strange'' weak value (STWV). 
The Hardy experiment, shown in Figure 1, consists of two overlapping interferometers, one containing
an electron, e-, and the other a positron, e+. The interferometers are precisely tuned
in such a way that if both e- and e+ are in the overlapping arms, they
will meet and annihilate one another. There are two detectors C and D in each interferometer,
one of which (D) can only be activated if there is an object in the overlapping
arm. The curious feature is that it is possible for both D's (i.e., D- for the
electron and D+ for the positron) to click and yet for the e-, e+ pair
not to annihilate one another (i.e., not to both be in the overlapping
arms). Note that this corresponds exactly to a failure
of the consistency condition (5),
for it contradicts the classically necessary idea that either the
electron (positron) is in the overlapping arm and the detector D-(D+)
can click {\it or} the electron (positron) is not in the overlapping arm
and the detector D-(D+) cannot click. That is, these two histories
{\it should} be mutually exclusive, but quantum mechanically they seem
not to be.
\vskip .5cm
\special{em:linewidth 0.4pt}
\unitlength 1.00mm
\linethickness{0.4pt}
\begin{picture}(137.00,133.67)
\emline{28.00}{50.33}{1}{28.00}{114.33}{2}
\emline{28.00}{114.33}{3}{79.33}{114.33}{4}
\emline{79.33}{114.33}{5}{79.33}{64.67}{6}
\emline{79.33}{64.67}{7}{28.00}{64.67}{8}
\emline{67.66}{80.33}{9}{120.00}{80.33}{10}
\emline{120.00}{80.33}{11}{120.00}{25.67}{12}
\emline{120.00}{25.67}{13}{67.33}{25.67}{14}
\emline{67.33}{25.67}{15}{67.33}{80.33}{16}
\emline{67.33}{26.00}{17}{52.00}{26.00}{18}
\emline{79.33}{114.33}{19}{79.33}{129.33}{20}
\emline{79.33}{114.33}{21}{93.66}{114.33}{22}
\emline{120.00}{80.33}{23}{120.00}{94.67}{24}
\emline{120.00}{94.67}{25}{120.00}{94.67}{26}
\emline{120.00}{80.33}{27}{132.66}{80.33}{28}
\put(28.00,44.33){\makebox(0,0)[cc]{p}}
\put(44.66,26.00){\makebox(0,0)[cc]{e}}
\emline{25.33}{62.67}{29}{30.33}{67.33}{30}
\emline{25.33}{112.33}{31}{30.66}{116.67}{32}
\emline{76.66}{112.00}{33}{82.33}{116.33}{34}
\emline{75.66}{62.00}{35}{83.00}{67.67}{36}
\emline{64.33}{78.00}{37}{70.33}{83.00}{38}
\emline{64.00}{22.33}{39}{70.66}{29.00}{40}
\emline{116.66}{77.33}{41}{123.66}{83.33}{42}
\emline{116.33}{22.00}{43}{124.00}{28.67}{44}
\put(79.33,133.67){\makebox(0,0)[cc]{$C+$}}
\put(100.33,114.33){\makebox(0,0)[cc]{$D+$}}
\put(120.33,99.67){\makebox(0,0)[cc]{$C-$}}
\put(137.00,80.67){\makebox(0,0)[cc]{$D-$}}
\put(18.00,117.67){\makebox(0,0)[cc]{${|NO\rangle}_p$}}
\put(109.33,15.33){\makebox(0,0)[cc]{${|NO\rangle}_e$}}
\put(45.66,69.33){\makebox(0,0)[cc]{${|O\rangle}_p$}}
\put(62.33,49.00){\makebox(0,0)[cc]{${|O\rangle}_e$}}
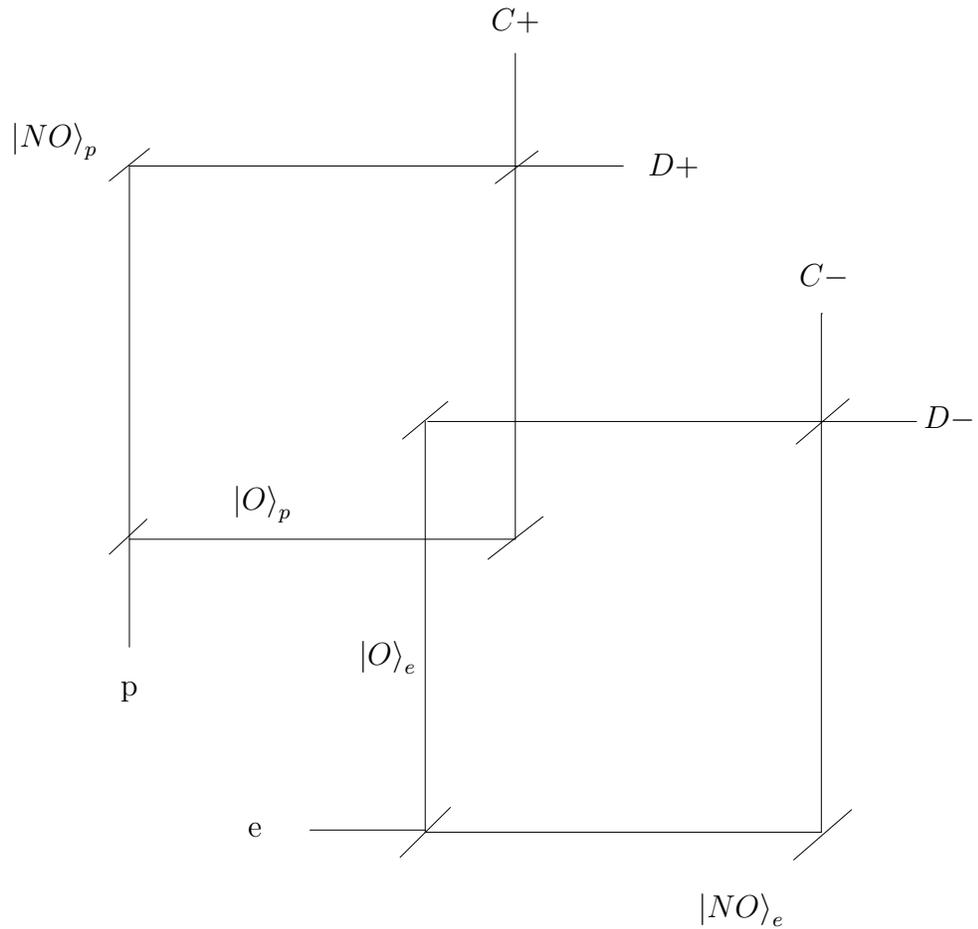
\put(65.00,3.00){\makebox(0,0)[cc]{Figure 1. Hardy's interferometer experiment.}}
\end{picture}

\vskip .5cm
Aharonov et al. (2001) show that the Hardy setup
gives rise to a ``strange'' weak value of $-1$ for the observable corresponding
to e+ and e- both being in the non-overlapping arms.

Let us confirm that this weak value indicates an inconsistent
family of histories. First some terminology:

The states are ${|O\rangle}$ and ${|NO\rangle}$
for a particle in the overlapping or non-overlapping arm, respectively;
a subscript of p or e denotes the positron or the electron.  
The two-particle states are constructed as direct products of
the single-particle states.

The clicking of detectors C correspond to the states 
${1\over \sqrt 2} ( |O\rangle + |NO\rangle )$, and the
clicking of detectors D correspond to the states
${1\over \sqrt 2} ( |O\rangle - |NO\rangle )$.

After the the electron and positron
pass the first beam splitter, they are in the state

$${1\over \sqrt 2} \biggl( {|O\rangle}_p + {|NO\rangle}_p \biggr) 
\otimes {1\over \sqrt 2} \biggl( {|O\rangle}_e + {|NO\rangle}_e \biggr) \eqno(16)$$

Aharonov et al want to look at the case in which the
electron and positron are not annihilated and yet the
detectors D+ and D- do click (which is the one which is puzzling).
Thus they project out the part of state (16) corresponding
to annihilation, which is ${|O\rangle}_p{|O\rangle}_e$,
 and use what remains as the preselection state:

$$|\psi\rangle = {1\over \sqrt 3} \biggl( {|NO\rangle}_p{|NO\rangle}_e
 + {|NO\rangle}_p{|O\rangle}_e + {|O\rangle}_p{|NO\rangle}_e \biggr)\eqno(17)$$

They use the post-selection state corresponding to the
clicking of the detectors D+ and D-, which is:

$$|\phi\rangle = {1\over 2} \biggl( {|NO\rangle}_p{|NO\rangle}_e
 + {|O\rangle}_p{|O\rangle}_e - {|NO\rangle}_p{|O\rangle}_e
 - {|O\rangle}_p{|NO\rangle}_e \biggr)\eqno(18)$$

\vskip .5cm
Now, using a convenient shorthand for the pair-occupation basis, viz.

$$|1\rangle \equiv {|NO\rangle}_p{|NO\rangle}_e$$

$$|2\rangle \equiv {|O\rangle}_p{|O\rangle}_e$$

$$|3\rangle \equiv {|NO\rangle}_p{|O\rangle}_e$$

$$|4\rangle \equiv {|O\rangle}_p{|NO\rangle}_e
\eqno(19a,b,c,d)$$
\vskip .5cm

The pre- and post-selection states respectively are:

$$|\psi\rangle = {1\over \sqrt 3} \biggl( |1\rangle + |3\rangle + 
|4\rangle \biggr)\eqno(20)$$

$$|\phi\rangle = {1\over 2} \biggl( |1\rangle + |2\rangle - |3\rangle - |4\rangle
 \biggr)\eqno(21)$$
\vskip .5cm
Now, the ``strange'' weak value of $-1$ arises for the pair in the non-overlapping arms,
state $|1\rangle$ in this notation. It is easily verified that the consistency
condition fails in the second, dramatic sense described in section 2 (where $E = |1\rangle\langle 1|$ and 
$ E' = |2\rangle\langle 2| + |3\rangle\langle 3| + |4\rangle\langle 4|$):

$$|\langle \phi | \psi\rangle |^2  {\langle E\rangle}_w {\langle E'\rangle}^*_w = 
{1\over 12} \biggl( -1 \biggr) \lbrack 0 + 1 + 1\rbrack = -{1\over 6} < 0 \eqno(22)$$

To arrive at an interpretation of the STWV of $-1$ as the answer to
a counterfactual question, Aharonov et al use what
they term ``Rule (a)'': the fact that
if an outcome of a measurement of an observable $A$ is known
with certainty to be the eigenvalue $a$, then the weak value ${\langle A\rangle}_w$ is equal to
that particular eigenvalue. They claim
that this allows one to combine two or more such outcomes using the 
additivity of the weak values (see equation (2)) to arrive
at apparently equally certain counterfactual inferences 
about the resultant, sum observable.

Thus they argue that using the additivity of weak values,
i.e., the fact that

$$ {\langle A + B\rangle}_w = {\langle A \rangle}_w + {\langle B \rangle}_w, 
  \eqno(2)$$

\noindent one
can combine sharp weak values (SWV) to obtain the strange
weak value (STWV), in the following way:

Given that 

$$|1\rangle\langle 1| + |2\rangle\langle 2| + |3\rangle\langle 3| +
|4\rangle\langle 4| = 1, \eqno(23)$$

\noindent (since equations (19) consitute a basis for the Hilbert Space) and
that 

$$ |1\rangle\langle 1| = 1 -  |2\rangle\langle 2|
-  |3\rangle\langle 3| -  |4\rangle\langle 4|, \eqno(24)$$

\noindent therefore the weak value of $|1\rangle\langle 1|$
is given by

$${\langle |1\rangle\langle 1|\rangle}_w = 1 - {\langle |2\rangle\langle 2|\rangle}_w
- {\langle |3\rangle\langle 3|\rangle}_w - {\langle |4\rangle\langle 4|\rangle}_w =
1- 0 - 1- 1 = -1.       \eqno(25)$$

They then suggest a physical (albeit counterintuitive) interpretation of
this calculation in which there is ``minus one electron-positron pair''
in the non-overlapping arms, and relate this interpration to counterfactual
statements about the whereabouts of the particles. This possible 
connection of weak values to counterfactual questions is considered
in detail in sections 6 and 7.

5. Time Symmetry Considerations

In this section I consider an important component of many discussions
of weak values: the time symmetry of pre- and post-selection. The idea
is that there is not only a state vector $|\psi\rangle$ propagating forward in time
from the pre-selection, but also a time-reversed adjoint vector $\langle \phi|$
propagating backward in time. Since physical laws are fundamentally
time-symmetric,{\footnote{\normalsize There are, of course,
certain phenomena such as the decay of the neutral K meson which
suggest some empirical deviation from strict time-reversal invariance.
The time symmetry considerations in this paper are restricted to the formalism of nonrelativistic
quantum theory, in which all laws are time-symmetric.}
there is nothing {\it a priori} wrong
with this assumption. Let us see what it adds to the analysis
in section 3, of the three-box experiment. 

Consider equation (14) which expresses the state of the combined system
at the intervening time $t_1$. If we adopt the idea that the backward-propagating
post-selection state $\langle \phi|$ has the same status as the forward propagating state
$|\psi\rangle$, we have to consider that the system somehow ``knows'' about 
its future post-selection, or bears that imprint, just as much as it
bears the imprint of its preselection. (For arguments in favor of this
approach, cf. Aharonov and Vaidman (1990) or Price (1996).) So let us consider a given
particle as ``fated'' to be post-selected in state $\postone$. We might therefore
conclude that the second term in (14) is not applicable 
(notice that this departs from standard quantum mechanics)
and that therefore one can consider the combined system to be ontologically
describable by (15) rather than (14) at $t_1$. In that case the 
apparatus is describable by the superposition
in (15) which reflects the weak value. This conclusion differs from that of Busch (1988),
who assumes time asymmetry, with causality moving only from the
past to the future. 

However, even if we consider the
total system as described by (15) at $t_1$, rather than by (14), what this
tells us is that the particle is described by 
its pre- and post-selection states and that the apparatus is in
a superposition of pointer states, i.e., it has an indeterminate
pointer value. The weak value of $-1$, which arises
from an averaging procedure over many runs, cannot be considered as 
applicable to either the particle or the apparatus in any particular
run of the experiment.

6. ``Strange'' weak values and the ABL rule

Before considering the meaning of the strange weak value of $-1$, first let us 
compute what might seem to be the corresponding probability of finding the
particle in box C: the value given by the much-discussed ABL rule (Aharonov,
Bergmann, and Lebowitz [1964]). The ABL rule gives the probability of
an outcome of an observable {\it actually (sharply) measured} at time $t_1$ given 
known pre- and post-selected states at $t_0$ and $t_2$ respectively.

In this case we need to use the form of the ABL rule appropriate
for degenerate operators, first presented in Aharonov and Vaidman (1991).
First some notation: let the projection
operator on the space of eigenstates corresponding to the value
$x$ be denoted by $P_x$. In this case, the two possible eigenvalues
are $c$ or $c'$ where the latter indicates ``not in box C.'' 
So the two projection operators will be $P_c = |c\rangle\langle c|$
and $P_{c'} = 1-P_c = |a\rangle\langle a| + |b\rangle\langle b|$. 
Then the ABL probability for outcome ``particle in box C,''
given the above pre- and post-selected states, and provided
C was actually opened, is:

$$P(c)= {|\langle \phi|P_c|\psi\rangle|^2
\over \sum_i |\langle \phi|P_i|\psi\rangle|^2} =
{|\langle \phi|P_c|\psi\rangle|^2
\over {|\langle \phi|P_c|\psi\rangle|^2 + |\langle \phi|P_{c'}|\psi\rangle|^2}} =
{1\over 5} \eqno(26)$$

Therefore, it is uncontroversially correct to say of any given particle in such
an ensemble that if box C was {\it in fact} opened, that particle had a 
probability of \ $1\over 5$ \ of being in the box. 
Note that this value corresponds to a non-counterfactual
situation since the measurement has occurred; i.e, the particle has
been disturbed, and is described at $t_1$ by an ignorance-type mixed state
in the eigenspace of the C observable. 
In contrast, the weak measurement is supposed to leave the particle
undisturbed, so it is still in a pure state. 

The ABL rule can also be expressed
in terms of the relevant weak values,
${\langle P_c\rangle}_w = {\langle \phi|P_c|\psi\rangle\over \langle \phi|\psi\rangle}$
and ${\langle P_{c'}\rangle}_w = {\langle \phi|P_{c'}|\psi\rangle\over 
\langle \phi|\psi\rangle}$:

$$P(c)_{ABL}= {|\langle \phi|P_c|\psi\rangle|^2
\over {|\langle \phi|P_c|\psi\rangle|^2 + 
|\langle \phi|P_{c'}|\psi\rangle|^2}}$$

$$ = {|{\langle P_c\rangle}_w|^2 |\langle \phi|\psi\rangle|^2 \over 
{|{\langle P_c\rangle}_w|^2 |\langle \phi|\psi\rangle|^2 + 
|{\langle P_{c'}\rangle}_w|^2 |\langle \phi|\psi\rangle|^2}}
={ |{\langle P_c\rangle}_w|^2 \over 
{|{\langle P_c\rangle}_w|^2  + |{\langle P_{c'}\rangle}_w|^2 } }\eqno(27)$$

Now, Using the fact that $P_{c'} = 1-P_c$ and the additivity of weak values,
we find

$$P(c)_{ABL} ={ |{\langle P_c\rangle}_w|^2 \over 
{|{\langle P_c\rangle}_w|^2  + |1-{\langle P_c\rangle}_w|^2 } }\eqno(28)$$

Obviously, if we substitute the value $-1$ for ${\langle P_c\rangle}_w$,
(28) still gives $1\over 5$ for the probability of finding the
particle in box C. This result presents a problem for claims
that the ABL rule always gives valid results for counterfactual (or weak)
 measurements on pre- and post-selected systems regardless of whether 
the history in question belongs to a consistent family. For
 (28), which 
gives the ABL probability associated with the relevant weak value,
tells us that the particle is in box
C 20\% of the time, a perfectly normal figure that would seem to have
nothing whatever to do with a strange weak value of $-1$.  So there seems to be an inconsistency arising
between the weak value itself and the counterfactual ABL probability associated with that
weak value. (This is not a problem for the case of an
{\it actual} $C$ measurement at $t_1$ since the particle has been disturbed
and its state projected into a classical probability space.) 

There is a different expression available that gives a 
weight
of an outcome conditional on other known
outcomes, and which avoids the above inconsistency.
It is obtained by extending
Luders' rule to the multiple time case, to obtain the following
weight of a history such as Y (3):

$$ W(Y)= Tr [D E F E]  \eqno (29)$$

From this one can obtain the ``conditional weight''

$$ W(E|D, F) =  {Tr [DEFE]\over Tr[DF]} .\eqno(30)$$

Note that this expression cannot in general be termed a ``probability''
because it does not always yield a value in the interval $(0,1)$.
This is because it is not restricted to consistent families
of histories. However, we can think of it as a pseudo-probability.
For a more detailed discussion of this expression, cf. Griffiths (1996),
Saunders (2001).

In the particular case at hand, we would then have

$$W(C|\psi, \phi) = {Tr(P_{\phi} P_C P_{\psi}P_C)\over {Tr(P_{\phi}P_{\psi})}}
 \eqno(31)$$

But note that (31) is just equal to the weak value squared, since

$${Tr(P_{\phi} P_C P_{\psi}P_C)\over {Tr(P_{\phi}P_{\psi})}} = 
{\langle \phi|P_c|\psi\rangle\over \langle \phi|\psi\rangle} 
{{\langle\psi|P_c|\phi\rangle}\over {\langle \psi|\phi\rangle} } 
= |{\langle P_c \rangle}_w|^2. \eqno(32)$$

Thus, using (32) instead of the ABL rule, we have the result $1$ for the conditional 
weight
of the particle's being in box C given that it has been pre- and
post-selected with no (or only a weak) measurement at time $t_1$.
This result, as bizarre as it is considering that the
weights associated with box A and
box B are also unity, has an appropriate relationship with the strange
weak value of $-1$ for the number of particles in box C at time $t_1$,
if we think of the weak value as being an ``amplitude'' corresponding
to the $C$ observable
conditional on the given pre- and post-selection. More will be
said in section 7 
on why the weak value should be thought of as an amplitude,
and the relevance of this point to counterfactual questions.

Again, notice that this inconsistency between the probability
given by the ABL rule and the weak value itself only arises
for weak values corresponding to inconsistent histories. 
Weak values corresponding to consistent histories are
$0$ or $1$, and substituting either of these into (28)
gives $0$ or $1$ respectively (since the right hand side is equal to
${\langle P_c\rangle}_w$ if and only if $|{\langle P_c\rangle}_w|^2 = 
{\langle P_c\rangle}_w$.)  This is because the ABL rule (28) and
 the conditional weight (30) are equivalent for 
histories belonging to consistent families, as shown in Kastner (1999b).
Indeed, for situations only involving a single consistent family,
(30) does function as a legitimate probability. Thus it is simply
a more general expression than (28).

7. Weak Values as Amplitudes, Not Expectation Values

Now, back to the idea of the weak value as an
amplitude, and what this implies for counterfactuals.
In discussions of the Three-Box and Hardy experiments, some authors
have suggested that weak values can serve as answers to counterfactual
questions for sets of measurements that cannot all be performed
on a given system. For example, Vaidman (1996) suggests that
the weak value of $-1$ for the particle in box C can be taken
as an answer to the question: ``If I had opened box C (instead
of A or B), would the particle have been there?''(and that
the $-1$ value indicates that the answer is not just ``no'' but
that there is somehow {\it less than zero particles}
in box C).  Similarly,
in Aharonov et al it is suggested that the weak value of
$-1$ for the electron-positron pair in the nonoverlapping
arms can serve as a (bizarre) answer to a counterfactual question
about the presence of the pair in that location when they have not 
actually been looked for there.

 In considering these claims it
seems relevant to recall that the conditional weight for a series of three 
events 
given a first and last event (equation 30-32) is
the absolute square of the weak value, not the weak value itself. 
This conditional weight functions as a standard probability when
the weak value corresponds to a consistent family of histories;
indeed, for these cases, (28) and (30-32) are equivalent.
It has already been shown that strange weak values (STWV) correspond
to inconsistent families in which the consistency condition fails
in a particularly extreme way, so one should not expect the ABL rule (28)
to apply in these cases. What is now applicable instead is the
more general conditional weight (30-32), despite the
fact that in these cases the conditional weight will not be
well-behaved as a probability. For, in keeping with the ``strangeness''
of these weak values, one should not expect to obtain a ``normal''
or well-behaved probability.

The fact that we must take the absolute square of the weak value
to obtain the appropriate probability or weight corresponding to that weak value,
coupled with the fact that the weak value can take
on negative and complex values, suggests that the weak value
should be thought of as an amplitude and not an expectation value
at all. In that case, it could give no physically relevant answer to
any question about the property of a system. That is, claims
such as ``there is -1 particle in box C'' based on a weak
value of the observable $C$ invokes a figure that is only an amplitude,
not a proper expectation value (which always gives a real result
within the range of eigenvalues).

The foregoing discussion pertains to Aharonov et al's ``rule (a),'' mentioned in
section 4, in the following way. This rule is derivable
from the fact that the outcomes which are certain are those
corresponding to weak values of either $0$ or $1$. 
Interpreting the weak value
as an amplitude as in (32), one obtains a corresponding 
probability\footnote{\normalsize In these cases, (32) functions
as a legitimate probability because these weak values
correspond to consistent families.} of
either zero or unity since the values $0$ and $1$ remain
the same when squared, hence those weak values
correspond to a certain outcome and, despite being
only amplitudes, can be taken
as physically meaningful. But when one combines such ``certain'' weak values
via additivity to obtain STWVs, one has combined mutually inconsistent
histories. The combining of inconsistent histories
is precisely what causes (32) to cease being a well-behaved
probability, which means that the probability space for
the possible outcomes corresponding to those STWVs is not well-defined. 
In turn, this means that the counterfactual status of such
outcomes cannot be well-defined either (nor, as mere amplitudes,
can STWVs give substantive physical information about the
system). 

The above argument is 
in accordance with arguments by Cohen (1995), 
Griffiths (1999, 2002, sec. 22.5), that combining
outcomes belonging to inconsistent histories 
does not yield valid counterfactual inferences.

The puzzling feature of all this is the following: Weak values {\do} obey 
mathematical additivity,
yet when we ``add'' them we get bizarre results. The puzzle is 
(at least partially) solved
when we consider weak values as amplitudes, since quantum mechanical amplitudes
are not expected to yield real properties to begin with.

8. Conclusion

The consistency criterion of Griffiths (cf. 1996, 2002) is expressed in
terms of weak values of projection operators and it is shown that  
``strange'' weak values (those which fall outside the range of
eigenvalues)
correspond to an inconsistent family of histories in which the consistency criterion
fails in a particularly ``bad'' way (i.e., the right hand side is
negative or complex).
It is also shown that using the ABL rule to obtain probabilities corresponding
to strange weak values, such as $-1$, gives inconsistent results for the
case of weak or counterfactual measurements.
Using instead an expression based on the multiple-time Luders' rule, 
the conditional weight,
which is equal to the square of the weak value, gives
more consistent results for such measurements. 
It is argued that weak values should be thought of as amplitudes,
and as such cannot be expected to always give physically meaningful answers to counterfactual
questions about the values of observables in the context
of pre- and post-selection.

In addition it is
argued that assuming a complete lack of disturbance of systems during weak measurement
together with a reverse causality postulate results
in the conclusion that the apparatus superposition reflecting the weak value can be considered 
applicable at the time $t_1$ of the measurement.

It is certainly true that STWVs raise interesting questions, and that post-selection brings with it some
surprising and useful features (cf. Mermin (1997), Bub (2001)). 
It is my hope to have suggested a possible avenue to solving
some of the puzzles posed by ``strange'' weak values.

\vskip 1cm
Acknowledgements

The author wishes to thank J. Bub, R. B. Griffiths, J. Finkelstein,
and an anonymous referee
for valuable comments and suggestions. 

This research is supported in part by grant no. SES-0115185
of the National Science Foundation.

\newpage
References

\noindent Y. Aharonov, A. Botero, S. Popescu, B. Reznik, J. Tollaksen (2001).
`Revisiting Hardy's Paradox: Counterfactual Statements, Real Measurements,
Entanglement and Weak Values,.'' e-print, quant-ph/0104062.\newline
Y. Aharonov and L. Vaidman (1990). `Properties of a quantum system during
the time interval between two measurements,' {\it Physical Review A 41},
11-20. \newline
Y. Aharonov and L. Vaidman (1991). `Complete Description
of a Quantum System at a Given Time,'
{\it Journal of Physics A 24}, 2315-28.\newline
J. Bub (2000). `Secure Key Distribution via Pre- and Post-Selected
Quantum States, e-print quant-ph/0006086, forthcoming in {\it Phys. Rev. A}.\newline
J. Bub (1997). {\it Interpreting the Quantum World}. Cambridge University
Press.\newline
P. Busch (1988). `Surprising Features of Unsharp Quantum Measurements,'
{\it Phys. Lett. A 130}, 323-329.\newline
O. Cohen (1995). `Pre- and postselected quantum systems, 
counterfactual 
measurements, and consistent histories, '
{\it Physical Review A 51}, 4373-4380.\newline
R. P. Feynman (1987). `Negative Probabilities,' in Hiley and Peat, eds. (1987).
{\it Quantum Implications: Essays in Honor of David Bohm}, pp.235-248, London: Routledge.\newline
R. B. Griffiths (1996). ``Consistent Histories and Quantum Reasoning,'' 
{\it Phys. Rev. A 54}, 2759.\newline
R. B. Griffiths (1999). ``Consistent Quantum Counterfactuals,'' 
{\it Phys.Rev. A60}, 5-8.\newline
R. B. Griffiths (2002). {\it Consistent Quantum Theory}. Cambridge:
Cambridge University Press.\newline
L. Hardy (1992). {\it Phys. Rev. Lett. 68}, 2981.\newline
R. E. Kastner (1999a). `Time-Symmetrised Quantum Theory, Counterfactuals
and Advanced Action,' {\it Studies in History and Philosophy of Modern Physics 30},
237-259.\newline
R. E. Kastner (1999b). `The Three-Box Paradox and Other Reasons to Reject
the Counterfactual Usage of the ABL Rule,' {\it Foundations of Physics 29}, 851-863.\newline
N. D. Mermin (1997). 'How to Ascertain the Values of 
Every Member of a Set of Observables That Cannot All 
Have Values,' in R. S. Cohen et al. (eds), 
{\it Potentiality, Entanglement and Passion-at-a-Distance}. 
Kluwer Academic Publishers, 149-157.\newline
H. Price (1996). {\it Time's Arrow and Archimedes' Point}, Oxford: Oxford 
University Press.\newline
S. Saunders (2001). ``Space-Time and Probability,'' in {\it Chance in 
Physics}, eds. J. Bricmont et al., Springer-Verlag.\newline
L. Vaidman (1993). pp. 406-417, `Elements of Reality and the Failure of the
Product Rule,' in {\it Symposium on the Foundations
of Modern Physics}, P. J. Lahti, P. Busch, and P. Mittelstaedt (eds.),
World Scientific.\newline
L. Vaidman (1996). `Weak-Measurement Elements of Reality,' 
{\it Foundations of Physics 26}, 895-906.\newline
J. von Neumann (1932). `Measurement and Reversibility', and `The Process
of Measurement,' in {\it Mathematical Foundations of Quantum Mechanics},
Princeton University Press, as reprinted in {\it Quantum Theory
and Measurement}, 1983, J. A. Wheeler and W. H. Zurek, eds., Princeton
University Press.\newline
E. Wigner (1932). {\it Phys. Rev. 40}, 749.

\end{document}